\documentclass[aps,prl,twocolumn,superscriptaddress,showpacs]{revtex4}

\usepackage{longtable}
\usepackage{graphicx}
\usepackage{dcolumn}
\usepackage{bm}
\usepackage{amsmath}
\usepackage[psamsfonts]{amssymb}

\newcommand{\STO}{SrTiO$_3$}
\newcommand{\LAO}{LaAlO$_3$}

\newcommand{\MR}{magnetoresistance}

\newcommand{\etal}{\emph{et al.}}
\newcommand{\TDEG}{two dimensional electron gas}

\begin{document}

\title{Shubnikov-de Haas oscillations in \STO/\LAO~interface}


\affiliation{Raymond and Beverly Sackler School of Physics
and Astronomy, Tel-Aviv University, Tel Aviv, 69978, Israel}

\author{M. Ben Shalom}
\author{A. Ron}
\author{A. Palevski}
\author{Y. Dagan}
\email[]{yodagan@post.tau.ac.il}


\date{\today}

\begin{abstract}
Quantum magnetic oscillations in \STO/\LAO~ interface are observed. The evolution of their frequency and amplitude at various gate voltages and temperatures is studied. The data are consistent with the Shubnikov de-Haas theory. The Hall resistivity $\rho_{xy}$ exhibits nonlinearity at low magnetic field. $\rho_{xy}$ is fitted assuming multiple carrier contributions. The comparison between the mobile carrier density inferred from the Hall data and the oscillation frequency suggests multiple valley and spin degeneracy. The small amplitude of the oscillations is discussed in the framework of the multiple band scenario.

\end{abstract}


\maketitle
The \TDEG~ (2DEG) formed at the interface between two insulating perovskites
is a subject of intense scientific interest.\cite{mannhartreview} The most widely studied interface has been the one created between \STO~ and \LAO.\cite{OhtomoHwang} At low temperatures this 2DEG has a
superconducting ground state, whose critical temperature can be
modified by an electric field effect.\cite{CavigiliaGating}The nature of the charge carriers and their
origin are still a matter of debate.\cite{Nakagawa_no_O_defects,
pentchevaPicketPRL, popovictheoryfortwodeg, PhysRevLett.98.196802,
WillmottPRL, OkamotoMillis, kalabukhovOVac, SalluzzoPRLEReconstruction}
The thickness of the conducting layer has been estimated to be of a few
nanometers by both transport measurements \cite{benshalom}\cite{Reyrend} and by using a conducting atomic force microscope (AFM). \cite{Copie2Dmetalic} From the AFM data analysis Copie \etal~ conclude that two types of charge carriers screen the local electric fields.
The \MR~ is strongly anisotropic \cite{benshalom} and effected by a gate dependent spin-orbit interaction.\cite{BenSHalomPRL, CavigliaSpinorbit} The $\rho_{xy}$ is non linear in magnetic field,\cite{seri:180410, BenSHalomPRL} and therefore it does not relate in a simple way to the number of charge carriers. The effective mass has been estimated using elipsometry\cite{Bernhard_elipsometry} to be 2.2 m$_e$ with m$_e$ the electron mass. The number of charge carriers, their effective mass and the effect of gate voltage on their mobility are a subject of vigorous research.
\par
Quantum oscillations in magnetic fields have been extensively used to study the electronic properties of metals semiconductors and correlated systems. In the standard theory of Shubnikov-de Haas (SdH) the amplitude of the oscillating part of the resistance is \cite{ShoenbergBook}
\begin{equation}
\Delta R=4 R_c R_T R_D Sin[2\pi(\frac{F}{B}-\frac{1}{2}) \pm \frac{\pi}{4}]
\end{equation}
Where $R_c$ is the non-oscillating part of the resistance. The oscillation amplitude decays with temperature as given by
\begin{equation}
R_T=\frac{2\pi^{2}m^\ast k_{B} T}{\hbar e B}/sinh(\frac{2\pi^{2}m^\ast k_{B} T}{\hbar e B})
\end{equation}
with $m^\ast$ the quasiparticle effective mass. The Dingle factor is
\begin{equation}
R_D=exp(-\frac{\pi}{\omega_{c}\tau_D})
\end{equation}
with $\omega_c=\frac{eB}{m^{\ast}}$ and $\tau_D$ the Dingle scattering time, which is related to the Dingle temperature by $k_B T_D= \hbar / 2\pi \tau_D$. This factor determines the decay of the amplitude of the oscillations as the field decreases. Increasing the magnetic field, lowering the temperatures and decreasing the scattering rate should all lead to increasing oscillations amplitude.
\par
In this letter we report longitudinal and Hall resistance measurements at low temperatures and intense magnetic fields of up to 31.5 T. The high magnetic field enables us to detect quantum oscillations in the longitudinal resistivity, as well as nonlinearities in the Hall resistance. The data are consistent with the SdH theory, enabling us to extract the high mobility carrier concentration, their effective mass, and the scattering rate. In addition, we fit the Hall data over the entire field range assuming high and low mobility carriers. The obtained mobile carrier concentration is much higher than the one inferred from the SdH frequency. This suggests multiple valley and spin degeneracy. We find that the carrier concentration and the scattering rate of the high mobility band obtained from the transport agree with the SdH analysis within a factor of three. However, the amplitude of the oscillations is unexpectedly small.

\begin{figure}
\includegraphics[width=1\hsize]{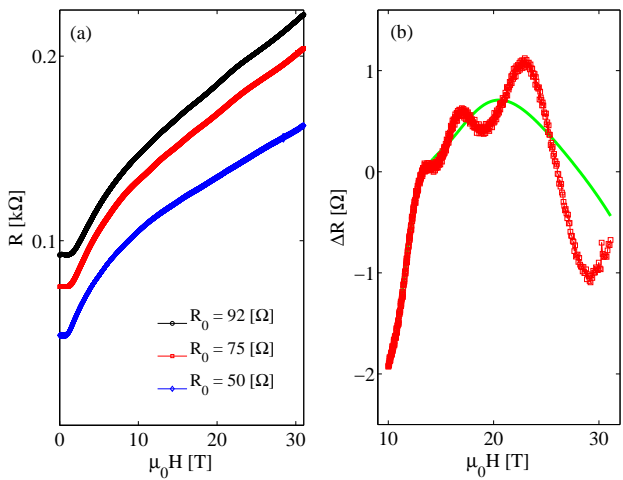}
\caption {(Color online) a. Sheet resistance as a function of magnetic field at 0.4 K. The zero field resistances is tuned by gate voltage. b. Squares: The curve for $R_0=75\Omega$ after a subtracting of a linear fit. The solid line is a polynomial background used in the data analysis.\label{fig1}}
\end{figure}

\par 
We use a sample with 15 unit cells of \LAO, deposited by pulsed laser on a TiO$_2$-terminated \STO (100) substrate. Deposition conditions are similar to Ref.\cite{BenSHalomPRL}. A $60 \times 120 \mu m^2$ Hall bar was patterned using the procedure of Schneider \etal, \cite{schneiderLithography} allowing a standard four-probe resistivity and Hall measurements. A gold layer was evaporated and used as a bottom gate when biased relative to the 2DEG. Aluminum or gold contacts were sputtered after drilling holes through the \LAO~ over-layer using an ion miller. The gate voltage was first set to $+200$ V. It was then used to control the state of the sample, i.e. the carrier concentration and the corresponding zero field resistivity $R_0$. We present the analysis as a function of $R_0$ and not the gate voltage since the latter may change from one cycle to the other.
\par
Figure \ref{fig1}a. presents the sheet resistance at base temperature (400mK) for 3 values of $R_0$ (controlled by the gate). As previously reported \cite{benshalom} a strong, positive \MR~ is observed. Fig.\ref{fig1}b. demonstrates the data of $R_0 = 75 \Omega$ (red squares) after subtraction of a straight line for $H > 10$ T. Conspicuous magnetic oscillations are observed.
\par
\begin{figure}
\includegraphics[width=1\hsize]{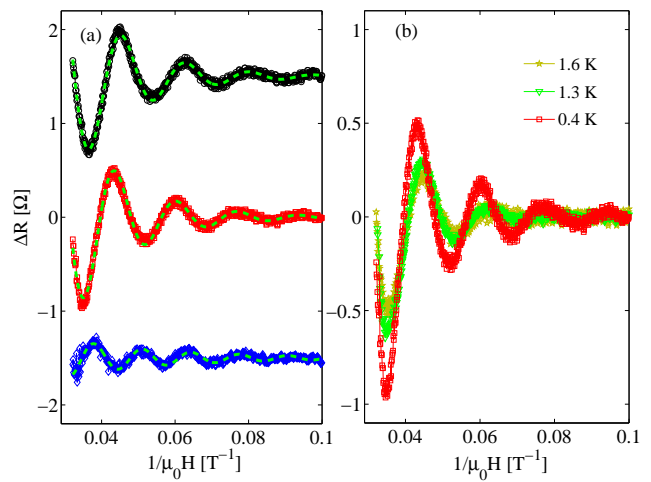}
\caption {(Color online) a. The data in fig.\ref{fig1} after subtraction of a polynomial background for the various $R_0$ The data are shifted for clarity by $1.5\Omega$. from top to bottom: $R_0=92, 75, 50\Omega$. The dashed lines are model fitting using equation(1). b. The data for $R_0=75\Omega$ for various temperatures after subtraction of a polynomial background. The value of $R_0$ is constant for this temperature range. \label{fig2}}
\end{figure}

Figure \ref{fig2}a. depicts $\Delta R$ versus $1/ \mu_0 H$ for the various $R_0$ after subtracting their polynomial background.  We made sure that this background is smooth and contains no oscillations in the field range under study. The background for $R_0 = 75 \Omega$ is shown in fig.\ref{fig1}b.(solid green line) .

Figure \ref{fig2}b. presents the same measurement for $R_0=75\Omega$ at three different temperatures. From the temperature dependence of the amplitude, using equation (2) we can extract $m^\ast$. At a constant field the oscillation amplitude depends solely on $m^\ast$ if we assume that $\tau$ is constant at this temperature range. This is consistent with the temperature independent resistance observed below 2 K (not shown). From the analysis we find $m^{\ast}=2.1 \pm 0.4 m_e$. The main errors stem from the uncertainty in temperature that may vary during the field scan. The complicated shape of the background introduces another source of uncertainty to the amplitude. Yet, since this background is smooth and temperature independent, its contribution to the uncertainty is minor. Moreover, upon using a band-pass filter instead of the polynomial subtraction we obtained the same effective mass.
\par
The dashed lines in Figure \ref{fig2}a. are the theoretical curves using Equation (1) leaving $R_c$ as a free parameter. This assumption will be discussed later. We find a good match for all values of $R_0$ and all temperatures when $R_0 = 75 \Omega$. For this gate voltage we find $T_D = 2\pm 0.4 K$. If we assume constant $m^\ast$ we estimate $T_D = 2.1$ and $1.5  K$ for $R_0 = 92$ and $50 \Omega$ respectively. It is possible, however, that the effective mass varies with carrier concentration.
\par
The various $R_0$ exhibit different frequencies in 1/H: 76, 60, 57.5 Tesla for $R_0$ = 50, 75, and 92 $\Omega$ respectively. The SdH frequency and the area of the Fermi surface are related by Onsager relation \cite{ShoenbergBook} $F=\frac{\hbar} {2 \pi e} A(\epsilon_F)$, where $A(\epsilon_F)$ is the area in momentum space of a closed orbit at the Fermi level. The \MR~ was measured while applying the field parallel to the current, no oscillations were observed (Not shown). This is consistent with a two-dimensional (2D) Fermi surface. According to Luttinger's theorem, $A(\epsilon_F)$ is directly related to $n_{2D}$ by: $n_{2D}=N_vN_seF/h$, where $N_v$ and $N_s$ are the valley and spin degeneracies, respectively. Ignoring any degeneracy, the obtained frequencies correspond to carrier densities of the order of $10^{12}$ cm$^{-2}$. This is much lower than the numbers obtained from the Hall coefficient or from those predicted theoretically.\cite{pentchevaPicketPRL} We attempt to reconcile this puzzle by a more careful analysis of the Hall resistivity, which is sensitive to the existence of two types of charge carriers (or more). In particular to low mobility carriers which do not contribute to the SdH effect.
\par

\begin{figure}
\includegraphics[width=1\hsize, height=0.75\hsize]{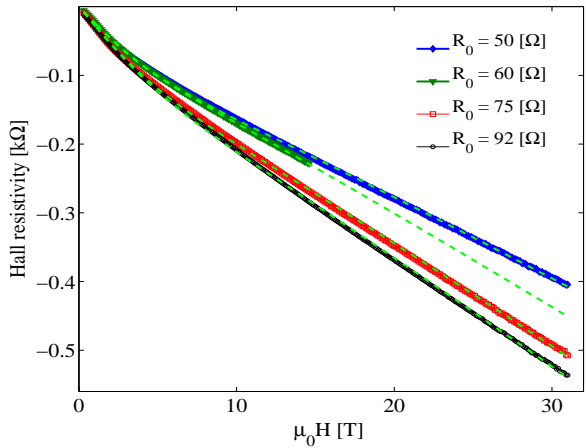}
\caption {(Color online) Hall resistivity as a function of magnetic field for the various $R_0$. The dashed lines are model fittings using equation(4). \label{fig3}}
\end{figure}

\begin{figure}
\includegraphics[width=1\hsize]{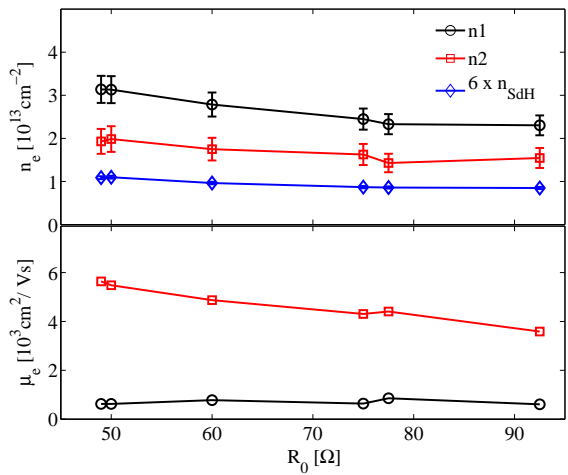}
\caption {(Color online) Upper panel: The carrier densities for the various $R_0$ inferred from the analysis of the Hall resistivity $n_1$ and $n_2$ correspond to the low and high mobility charge carrier density. The blue diamonds are the carrier densities obtained from the SdH frequency assuming 6 fold degeneracy. Lower panel: The mobilities of the two types of carriers inferred from the Hall analysis.\label{twoban}}
\end{figure}
Fig.\ref{fig3} shows the Hall resistivity measured for various gate voltages and sample resistances. A non-linear regime is observed at low magnetic fields. $\rho_{xy}$ is fitted using a two-band model. In this model the field dependence of $\rho_{xy}$ is given by
\begin{equation}
\rho_{xy}=\frac{\sigma_1^2R_1+\sigma_2^2R_2+\sigma_1^2\sigma_2^2R_1R_2(R_1+R_2)B^2}{(\sigma_1+\sigma_2)^2+\sigma_1^2\sigma_2^2(R_1+R_2)^2B^2}B
\end{equation}
with $R_i$ and $\sigma_i$ the Hall coefficient and conductivity of the $i^{th}$ type of carrier. The zero field resistance is $\frac{\rho_1 \rho_2}{\rho_1 + \rho_2}$. In fig.\ref{fig3} we show the fits to $\rho_{xy}$ for the entire field range using Equation (4) and the measured zero field resistance. From the fit we can extract the concentrations and mobilities of both types of carriers. Sometimes, the same model can also be used for fitting the \MR. However, the shape of the \MR~ is rather complicated and includes effects that cannot be described by a simple orbital model.\cite{benshalom} We therefore used only the zero field resistivity as a constraint for the fitting.
\par
The results of such fitting of the Hall resistivity are summarized in Figure \ref{twoban}. The upper panel presents the obtained carrier concentrations for the various $R_0$. The concentration $n_1$ of the low mobility charge carriers is about 50 percents larger than $n_2$, the concentration of the high mobility ones. The respective mobilities $\mu_1$ and $\mu_2$ are shown in the lower panel, were it can be seen that $\mu_2$ is 6-9 times larger. This observation explains the absence of oscillations corresponding to majority carrier concentration $n_1$. It seems that both bands respond to the gate variation.
\par
In addition, we present the carrier concentration $n_{SdH}$ obtained from the SdH analysis multiplied by six. The six-fold degeneracy is reasonable if one assumes a valley degeneracy of 3 and spin $1/2$. It is needed to better match $n_2$ and $n_{SdH}$. We have previously proposed that valley degeneracy of 3 is required to reconcile the root mean square amplitude and the width of the universal quantum fluctuations.\cite{davidpaper} The three-fold valley degeneracy is consistent with SdH measurements on SrTiO$_3$.\cite{STO_SdH}
The valley degeneracy picture is probably oversimplified. Furthermore, there is still a factor of two discrepancy between $n_{SdH}$ and $n_2$ which cannot be reconciled.
\par
The obtained mobilities exhibit different behavior with $R_0$ or gate voltage. While the mobility of the "slow" charge carriers, $\mu_1$ remains almost constant, $\mu_2$  increases with the total number of carriers. This is consistent with the observation of Bell \etal..\cite{Hwangmobility}
\par
We note that the phase of the SdH oscillations is determined by equation (1). This phase is calculated assuming parabolic bands.\cite{ShoenbergBook} Our data matches equation (1) including this phase for $R_0 = 75, 92 \Omega$. No other frequencies can be detected. It seems, however, that for $R_0= 50\Omega$ there is a phase shift of $\pi/2$ (minus sign is used in Equation(1)).
\par
All scattering processes contribute to $\tau_D$ while only backscattering enters the transport scattering time $\tau_t$. For $R_0=75 \Omega$, $\tau_D=6\times10^{-13}$ sec. From a simple calculation of the scattering time using the Hall mobility and the effective mass (found from the SdH analysis) we obtain $\tau_t=2\times10^{-12}$ sec. The factor of three difference is reasonable, taking into account the over simplified two-band analysis.
\par
The smearing of the Landau levels $k_BT_D\simeq0.2 meV$ is much smaller than their spacing $\hbar \omega_c = 1.65 meV$ (for 30 T). In addition the Landau level degeneracy is of the order $\phi/\phi_0=7\times10^{11}$, with $\phi_0$ being the flux quantum. This is not very far from the carrier concentration found from the naive SdH frequency analysis. Therefore, the amplitude of the SdH oscillations is expected to be very large. Yet, we observe a rather small effect $\Delta R/R \ll 1$. Moreover, for 400mK and 30T we get $R_T\simeq1$ and $R_D\simeq0.1$, therefore the oscillations are expected to be 10 percents of $R_c$. The amplitude of the oscillations $\simeq1 \Omega$ yields $R_c\simeq10 \Omega$, which is atleast by an order of magnitude smaller than the measured non-oscillating part of the resistance. I.e. $R_c\ll R_0$.
\par
The small amplitude of the SdH oscillations may be due to a three dimensional Fermi surface without Landau quantization in the parallel field orientation. This, however requires $R_L\gg d\gg \lambda_F$, with $R_L=\sqrt{\frac{h}{eB}}$ and  $\lambda_F=\frac{2\pi}{k_F}$. Using $n=10^{13} cm^{-2}$ carriers we find $\lambda_F\simeq10$ nm while $R_L\simeq12$ nm for B=30 T. This means that d, $\lambda_F$, and $R_L$ are all at the same order of magnitude and the inequality above is not easily obeyed. Another possibility is that the spin degeneracy is lifted, yet strong spin scattering results in smearing of the Zeeman dip. The resemblance of the data to the theoretical SdH curve cast doubt on this scenario.
\par
In summary, we measured longitudinal and transverse resistances in \STO/\LAO~ interfaces with various gate voltages and at various temperatures. The longitudinal resistance exhibits quantum oscillations with a single frequency. An analysis of the data according to the SdH theory yields the carrier density, effective mass, and scattering rate. The transverse resistance is fitted using a two types of charge carriers model. In order to reconcile these two measurements we had to invoke multiple valley and spin degeneracy. This yields an agreement within a factor of two in concentration and a factor of three in scattering rates. The small amplitude of the oscillations at high fields is still unclear.
\par
Our results suggest that separating out the low mobility band may result in a highly conducting oxide interface with gate tuning capabilities.

\par
During the preparation of the manuscript we became aware of another work reporting SdH oscillations in this system with a lower carrier concentration.\cite{CavigliaSdH} Their effective mass is somewhat smaller. This may be indicative of an effective mass variation with carrier concentration.
\par
This research was supported by the ISF under grant No. 1421/08, and by the
Wolfson Family Charitable Trust. A portion of this work was
performed at the National High Magnetic Field Laboratory, which is
supported by NSF Cooperative Agreement No. DMR-0654118, by the
State of Florida, and by the DOE.

\bibliographystyle{apsrev}
\bibliography{stolao}

\begin{thebibliography}{24}
\expandafter\ifx\csname natexlab\endcsname\relax\def\natexlab#1{#1}\fi
\expandafter\ifx\csname bibnamefont\endcsname\relax
  \def\bibnamefont#1{#1}\fi
\expandafter\ifx\csname bibfnamefont\endcsname\relax
  \def\bibfnamefont#1{#1}\fi
\expandafter\ifx\csname citenamefont\endcsname\relax
  \def\citenamefont#1{#1}\fi
\expandafter\ifx\csname url\endcsname\relax
  \def\url#1{\texttt{#1}}\fi
\expandafter\ifx\csname urlprefix\endcsname\relax\def\urlprefix{URL }\fi
\providecommand{\bibinfo}[2]{#2}
\providecommand{\eprint}[2][]{\url{#2}}

\bibitem[{\citenamefont{Mannhart and G.}(2010)}]{mannhartreview}
\bibinfo{author}{\bibfnamefont{J.}~\bibnamefont{Mannhart}} \bibnamefont{and}
  \bibinfo{author}{\bibfnamefont{S.~D.} \bibnamefont{G.}},
  \bibinfo{journal}{Science} \textbf{\bibinfo{volume}{327}},
  \bibinfo{pages}{1607} (\bibinfo{year}{2010}).

\bibitem[{\citenamefont{Ohtomo and Hwang}(2004)}]{OhtomoHwang}
\bibinfo{author}{\bibfnamefont{A.}~\bibnamefont{Ohtomo}} \bibnamefont{and}
  \bibinfo{author}{\bibfnamefont{H.~Y.} \bibnamefont{Hwang}},
  \bibinfo{journal}{Nature (London)} \textbf{\bibinfo{volume}{427}},
  \bibinfo{pages}{423} (\bibinfo{year}{2004}).

\bibitem[{\citenamefont{Caviglia~{\it et al.}}(2008)}]{CavigiliaGating}
\bibinfo{author}{\bibfnamefont{A.~D.} \bibnamefont{Caviglia~{\it et al.}}},
  \bibinfo{journal}{Nature (London)} \textbf{\bibinfo{volume}{456}},
  \bibinfo{pages}{624} (\bibinfo{year}{2008}).

\bibitem[{\citenamefont{Nakagawa et~al.}(2006)\citenamefont{Nakagawa, Hwang,
  and Muller}}]{Nakagawa_no_O_defects}
\bibinfo{author}{\bibfnamefont{N.}~\bibnamefont{Nakagawa}},
  \bibinfo{author}{\bibfnamefont{H.~Y.} \bibnamefont{Hwang}}, \bibnamefont{and}
  \bibinfo{author}{\bibfnamefont{D.~A.} \bibnamefont{Muller}},
  \bibinfo{journal}{Nature Mater.} \textbf{\bibinfo{volume}{5}},
  \bibinfo{pages}{204} (\bibinfo{year}{2006}).

\bibitem[{\citenamefont{Pentcheva and Pickett}(2009)}]{pentchevaPicketPRL}
\bibinfo{author}{\bibfnamefont{R.}~\bibnamefont{Pentcheva}} \bibnamefont{and}
  \bibinfo{author}{\bibfnamefont{W.~E.} \bibnamefont{Pickett}},
  \bibinfo{journal}{Phys. Rev. Lett.} \textbf{\bibinfo{volume}{102}},
  \bibinfo{eid}{107602} (\bibinfo{year}{2009}).

\bibitem[{\citenamefont{Popovi\'{c} et~al.}(2008)\citenamefont{Popovi\'{c},
  Satpathy, and Martin}}]{popovictheoryfortwodeg}
\bibinfo{author}{\bibfnamefont{Z.~S.} \bibnamefont{Popovi\'{c}}},
  \bibinfo{author}{\bibfnamefont{S.}~\bibnamefont{Satpathy}}, \bibnamefont{and}
  \bibinfo{author}{\bibfnamefont{R.~M.} \bibnamefont{Martin}},
  \bibinfo{journal}{Phys. Rev. Lett.} \textbf{\bibinfo{volume}{101}},
  \bibinfo{eid}{256801} (\bibinfo{year}{2008}).

\bibitem[{\citenamefont{Siemons et~al.}(2007)\citenamefont{Siemons, Koster,
  Yamamoto, Harrison, Lucovsky, Geballe, Blank, and
  Beasley}}]{PhysRevLett.98.196802}
\bibinfo{author}{\bibfnamefont{W.}~\bibnamefont{Siemons}},
  \bibinfo{author}{\bibfnamefont{G.}~\bibnamefont{Koster}},
  \bibinfo{author}{\bibfnamefont{H.}~\bibnamefont{Yamamoto}},
  \bibinfo{author}{\bibfnamefont{W.~A.} \bibnamefont{Harrison}},
  \bibinfo{author}{\bibfnamefont{G.}~\bibnamefont{Lucovsky}},
  \bibinfo{author}{\bibfnamefont{T.~H.} \bibnamefont{Geballe}},
  \bibinfo{author}{\bibfnamefont{D.~H.~A.} \bibnamefont{Blank}},
  \bibnamefont{and} \bibinfo{author}{\bibfnamefont{M.~R.}
  \bibnamefont{Beasley}}, \bibinfo{journal}{Phys. Rev. Lett.}
  \textbf{\bibinfo{volume}{98}}, \bibinfo{pages}{196802}
  (\bibinfo{year}{2007}).

\bibitem[{\citenamefont{Willmott et~al.}(2007)\citenamefont{Willmott, Pauli,
  Herger, Schlep\"utz, Martoccia, Patterson, Delley, Clarke, Kumah, Cionca
  et~al.}}]{WillmottPRL}
\bibinfo{author}{\bibfnamefont{P.~R.} \bibnamefont{Willmott}},
  \bibinfo{author}{\bibfnamefont{S.~A.} \bibnamefont{Pauli}},
  \bibinfo{author}{\bibfnamefont{R.}~\bibnamefont{Herger}},
  \bibinfo{author}{\bibfnamefont{C.~M.} \bibnamefont{Schlep\"utz}},
  \bibinfo{author}{\bibfnamefont{D.}~\bibnamefont{Martoccia}},
  \bibinfo{author}{\bibfnamefont{B.~D.} \bibnamefont{Patterson}},
  \bibinfo{author}{\bibfnamefont{B.}~\bibnamefont{Delley}},
  \bibinfo{author}{\bibfnamefont{R.}~\bibnamefont{Clarke}},
  \bibinfo{author}{\bibfnamefont{D.}~\bibnamefont{Kumah}},
  \bibinfo{author}{\bibfnamefont{C.}~\bibnamefont{Cionca}},
  \bibnamefont{et~al.}, \bibinfo{journal}{Phys. Rev. Lett.}
  \textbf{\bibinfo{volume}{99}}, \bibinfo{eid}{155502} (\bibinfo{year}{2007}).

\bibitem[{\citenamefont{Okamoto and Millis}(2004)}]{OkamotoMillis}
\bibinfo{author}{\bibfnamefont{S.}~\bibnamefont{Okamoto}} \bibnamefont{and}
  \bibinfo{author}{\bibfnamefont{A.~J.} \bibnamefont{Millis}},
  \bibinfo{journal}{Nature (London)} \textbf{\bibinfo{volume}{428}},
  \bibinfo{pages}{630} (\bibinfo{year}{2004}).

\bibitem[{\citenamefont{Kalabukhov et~al.}(2007)\citenamefont{Kalabukhov,
  Gunnarsson, B\"{o}rjesson, Olsson, Claeson, and Winkler}}]{kalabukhovOVac}
\bibinfo{author}{\bibfnamefont{A.}~\bibnamefont{Kalabukhov}},
  \bibinfo{author}{\bibfnamefont{R.}~\bibnamefont{Gunnarsson}},
  \bibinfo{author}{\bibfnamefont{J.}~\bibnamefont{B\"{o}rjesson}},
  \bibinfo{author}{\bibfnamefont{E.}~\bibnamefont{Olsson}},
  \bibinfo{author}{\bibfnamefont{T.}~\bibnamefont{Claeson}}, \bibnamefont{and}
  \bibinfo{author}{\bibfnamefont{D.}~\bibnamefont{Winkler}},
  \bibinfo{journal}{Phys. Rev. B} \textbf{\bibinfo{volume}{75}},
  \bibinfo{eid}{121404} (\bibinfo{year}{2007}).

\bibitem[{\citenamefont{Salluzzo et~al.}(2009)\citenamefont{Salluzzo, Cezar,
  Brookes, Bisogni, De~Luca, Richter, Thiel, Mannhart, Huijben, Brinkman
  et~al.}}]{SalluzzoPRLEReconstruction}
\bibinfo{author}{\bibfnamefont{M.}~\bibnamefont{Salluzzo}},
  \bibinfo{author}{\bibfnamefont{J.~C.} \bibnamefont{Cezar}},
  \bibinfo{author}{\bibfnamefont{N.~B.} \bibnamefont{Brookes}},
  \bibinfo{author}{\bibfnamefont{V.}~\bibnamefont{Bisogni}},
  \bibinfo{author}{\bibfnamefont{G.~M.} \bibnamefont{De~Luca}},
  \bibinfo{author}{\bibfnamefont{C.}~\bibnamefont{Richter}},
  \bibinfo{author}{\bibfnamefont{S.}~\bibnamefont{Thiel}},
  \bibinfo{author}{\bibfnamefont{J.}~\bibnamefont{Mannhart}},
  \bibinfo{author}{\bibfnamefont{M.}~\bibnamefont{Huijben}},
  \bibinfo{author}{\bibfnamefont{A.}~\bibnamefont{Brinkman}},
  \bibnamefont{et~al.}, \bibinfo{journal}{Phys. Rev. Lett.}
  \textbf{\bibinfo{volume}{102}}, \bibinfo{pages}{166804}
  (\bibinfo{year}{2009}).

\bibitem[{\citenamefont{Ben~Shalom et~al.}(2009)\citenamefont{Ben~Shalom, Tai,
  Lereah, Sachs, Levy, Rakhmilevitch, Palevski, and Dagan}}]{benshalom}
\bibinfo{author}{\bibfnamefont{M.}~\bibnamefont{Ben~Shalom}},
  \bibinfo{author}{\bibfnamefont{C.~W.} \bibnamefont{Tai}},
  \bibinfo{author}{\bibfnamefont{Y.}~\bibnamefont{Lereah}},
  \bibinfo{author}{\bibfnamefont{M.}~\bibnamefont{Sachs}},
  \bibinfo{author}{\bibfnamefont{E.}~\bibnamefont{Levy}},
  \bibinfo{author}{\bibfnamefont{D.}~\bibnamefont{Rakhmilevitch}},
  \bibinfo{author}{\bibfnamefont{A.}~\bibnamefont{Palevski}}, \bibnamefont{and}
  \bibinfo{author}{\bibfnamefont{Y.}~\bibnamefont{Dagan}},
  \bibinfo{journal}{Phys. Rev. B} \textbf{\bibinfo{volume}{80}},
  \bibinfo{eid}{140403} (pages~\bibinfo{numpages}{4}) (\bibinfo{year}{2009}).

\bibitem[{\citenamefont{Reyren~{\it et al.}}(2009)}]{Reyrend}
\bibinfo{author}{\bibfnamefont{N.}~\bibnamefont{Reyren~{\it et al.}}},
  \bibinfo{journal}{Appl. Phys. Lett.} \textbf{\bibinfo{volume}{94}},
  \bibinfo{eid}{112506} (\bibinfo{year}{2009}).

\bibitem[{\citenamefont{Copie et~al.}(2009)\citenamefont{Copie, Garcia,
  B\"odefeld, Carr\'et\'ero, Bibes, Herranz, Jacquet, Maurice, Vinter, Fusil
  et~al.}}]{Copie2Dmetalic}
\bibinfo{author}{\bibfnamefont{O.}~\bibnamefont{Copie}},
  \bibinfo{author}{\bibfnamefont{V.}~\bibnamefont{Garcia}},
  \bibinfo{author}{\bibfnamefont{C.}~\bibnamefont{B\"odefeld}},
  \bibinfo{author}{\bibfnamefont{C.}~\bibnamefont{Carr\'et\'ero}},
  \bibinfo{author}{\bibfnamefont{M.}~\bibnamefont{Bibes}},
  \bibinfo{author}{\bibfnamefont{G.}~\bibnamefont{Herranz}},
  \bibinfo{author}{\bibfnamefont{E.}~\bibnamefont{Jacquet}},
  \bibinfo{author}{\bibfnamefont{J.-L.} \bibnamefont{Maurice}},
  \bibinfo{author}{\bibfnamefont{B.}~\bibnamefont{Vinter}},
  \bibinfo{author}{\bibfnamefont{S.}~\bibnamefont{Fusil}},
  \bibnamefont{et~al.}, \bibinfo{journal}{Phys. Rev. Lett.}
  \textbf{\bibinfo{volume}{102}}, \bibinfo{pages}{216804}
  (\bibinfo{year}{2009}).

\bibitem[{\citenamefont{Ben~Shalom et~al.}(2010)\citenamefont{Ben~Shalom,
  Sachs, Rakhmilevitch, Palevski, and Dagan}}]{BenSHalomPRL}
\bibinfo{author}{\bibfnamefont{M.}~\bibnamefont{Ben~Shalom}},
  \bibinfo{author}{\bibfnamefont{M.}~\bibnamefont{Sachs}},
  \bibinfo{author}{\bibfnamefont{D.}~\bibnamefont{Rakhmilevitch}},
  \bibinfo{author}{\bibfnamefont{A.}~\bibnamefont{Palevski}}, \bibnamefont{and}
  \bibinfo{author}{\bibfnamefont{Y.}~\bibnamefont{Dagan}},
  \bibinfo{journal}{Phys. Rev. Lett.} \textbf{\bibinfo{volume}{104}},
  \bibinfo{pages}{126802} (\bibinfo{year}{2010}).

\bibitem[{\citenamefont{Caviglia et~al.}(2010)\citenamefont{Caviglia, Gabay,
  Gariglio, Reyren, Cancellieri, and Triscone}}]{CavigliaSpinorbit}
\bibinfo{author}{\bibfnamefont{A.~D.} \bibnamefont{Caviglia}},
  \bibinfo{author}{\bibfnamefont{M.}~\bibnamefont{Gabay}},
  \bibinfo{author}{\bibfnamefont{S.}~\bibnamefont{Gariglio}},
  \bibinfo{author}{\bibfnamefont{N.}~\bibnamefont{Reyren}},
  \bibinfo{author}{\bibfnamefont{C.}~\bibnamefont{Cancellieri}},
  \bibnamefont{and} \bibinfo{author}{\bibfnamefont{J.-M.}
  \bibnamefont{Triscone}}, \bibinfo{journal}{Phys. Rev. Lett.}
  \textbf{\bibinfo{volume}{104}}, \bibinfo{pages}{126803}
  (\bibinfo{year}{2010}).

\bibitem[{\citenamefont{Seri and Klein}(2009)}]{seri:180410}
\bibinfo{author}{\bibfnamefont{S.}~\bibnamefont{Seri}} \bibnamefont{and}
  \bibinfo{author}{\bibfnamefont{L.}~\bibnamefont{Klein}},
  \bibinfo{journal}{Phys. Rev. B} \textbf{\bibinfo{volume}{80}},
  \bibinfo{eid}{180410} (\bibinfo{year}{2009}).

\bibitem[{\citenamefont{Dubroka et~al.}(2010)\citenamefont{Dubroka, R\"ossle,
  Kim, Malik, Schultz, Thiel, Schneider, Mannhart, Herranz, Copie
  et~al.}}]{Bernhard_elipsometry}
\bibinfo{author}{\bibfnamefont{A.}~\bibnamefont{Dubroka}},
  \bibinfo{author}{\bibfnamefont{M.}~\bibnamefont{R\"ossle}},
  \bibinfo{author}{\bibfnamefont{K.~W.} \bibnamefont{Kim}},
  \bibinfo{author}{\bibfnamefont{V.~K.} \bibnamefont{Malik}},
  \bibinfo{author}{\bibfnamefont{L.}~\bibnamefont{Schultz}},
  \bibinfo{author}{\bibfnamefont{S.}~\bibnamefont{Thiel}},
  \bibinfo{author}{\bibfnamefont{C.~W.} \bibnamefont{Schneider}},
  \bibinfo{author}{\bibfnamefont{J.}~\bibnamefont{Mannhart}},
  \bibinfo{author}{\bibfnamefont{G.}~\bibnamefont{Herranz}},
  \bibinfo{author}{\bibfnamefont{O.}~\bibnamefont{Copie}},
  \bibnamefont{et~al.}, \bibinfo{journal}{Phys. Rev. Lett.}
  \textbf{\bibinfo{volume}{104}}, \bibinfo{pages}{156807}
  (\bibinfo{year}{2010}).

\bibitem[{\citenamefont{Shoenberg}(1984)}]{ShoenbergBook}
\bibinfo{author}{\bibfnamefont{D.}~\bibnamefont{Shoenberg}},
  \emph{\bibinfo{title}{Magnetic Oscillations in Metals}}
  (\bibinfo{publisher}{Cambridge University Press},
  \bibinfo{address}{Cambridge}, \bibinfo{year}{1984}).

\bibitem[{\citenamefont{Schneider et~al.}(2006)\citenamefont{Schneider, Thiel,
  Hammerl, Richter, and Mannhart}}]{schneiderLithography}
\bibinfo{author}{\bibfnamefont{C.~W.} \bibnamefont{Schneider}},
  \bibinfo{author}{\bibfnamefont{S.}~\bibnamefont{Thiel}},
  \bibinfo{author}{\bibfnamefont{G.}~\bibnamefont{Hammerl}},
  \bibinfo{author}{\bibfnamefont{C.}~\bibnamefont{Richter}}, \bibnamefont{and}
  \bibinfo{author}{\bibfnamefont{J.}~\bibnamefont{Mannhart}},
  \bibinfo{journal}{Applied Physics Letters} \textbf{\bibinfo{volume}{89}},
  \bibinfo{eid}{122101} (\bibinfo{year}{2006}).

\bibitem[{\citenamefont{Rakhmilevitch et~al.}()\citenamefont{Rakhmilevitch,
  Ben~Shalom, Eshkol, Tsukernik, Palevski, and Dagan}}]{davidpaper}
\bibinfo{author}{\bibfnamefont{D.}~\bibnamefont{Rakhmilevitch}},
  \bibinfo{author}{\bibfnamefont{M.}~\bibnamefont{Ben~Shalom}},
  \bibinfo{author}{\bibfnamefont{M.}~\bibnamefont{Eshkol}},
  \bibinfo{author}{\bibfnamefont{A.}~\bibnamefont{Tsukernik}},
  \bibinfo{author}{\bibfnamefont{A.}~\bibnamefont{Palevski}}, \bibnamefont{and}
  \bibinfo{author}{\bibfnamefont{Y.}~\bibnamefont{Dagan}} (????),
  \bibinfo{note}{submitted}.

\bibitem[{\citenamefont{Frederikse et~al.}(1967)\citenamefont{Frederikse,
  Hosler, Thurber, Babiskin, and Siebenmann}}]{STO_SdH}
\bibinfo{author}{\bibfnamefont{H.~P.~R.} \bibnamefont{Frederikse}},
  \bibinfo{author}{\bibfnamefont{W.~R.} \bibnamefont{Hosler}},
  \bibinfo{author}{\bibfnamefont{W.~R.} \bibnamefont{Thurber}},
  \bibinfo{author}{\bibfnamefont{J.}~\bibnamefont{Babiskin}}, \bibnamefont{and}
  \bibinfo{author}{\bibfnamefont{P.~G.} \bibnamefont{Siebenmann}},
  \bibinfo{journal}{Phys. Rev.} \textbf{\bibinfo{volume}{158}},
  \bibinfo{pages}{775} (\bibinfo{year}{1967}).

\bibitem[{\citenamefont{Bell et~al.}(2009)\citenamefont{Bell, Harashima,
  Kozuka, Kim, Kim, Hikita, and Hwang}}]{Hwangmobility}
\bibinfo{author}{\bibfnamefont{C.}~\bibnamefont{Bell}},
  \bibinfo{author}{\bibfnamefont{S.}~\bibnamefont{Harashima}},
  \bibinfo{author}{\bibfnamefont{Y.}~\bibnamefont{Kozuka}},
  \bibinfo{author}{\bibfnamefont{M.}~\bibnamefont{Kim}},
  \bibinfo{author}{\bibfnamefont{B.~G.} \bibnamefont{Kim}},
  \bibinfo{author}{\bibfnamefont{Y.}~\bibnamefont{Hikita}}, \bibnamefont{and}
  \bibinfo{author}{\bibfnamefont{H.~Y.} \bibnamefont{Hwang}},
  \bibinfo{journal}{Phys. Rev. Lett.} \textbf{\bibinfo{volume}{103}},
  \bibinfo{eid}{226802} (\bibinfo{year}{2009}).

\bibitem[{\citenamefont{Caviglia~{\it {et al.}}}()}]{CavigliaSdH}
\bibinfo{author}{\bibfnamefont{A.~D.} \bibnamefont{Caviglia~{\it {et al.}}}},
  \bibinfo{journal}{arXiv:1007.4941}  (????).

\end{thebibliography}
\end{document}